\documentclass[manuscript]{aastex}

\shorttitle{Collapsed Cores in Globular Clusters}
\shortauthors{Djorgovski et al.}

\begin{document}


\title{Evaporating Quantum Lukewarm Black Holes Final State From Back-Reaction Corrections
of Quantum Scalar Fields}


\author{H. Ghaffarnejad \altaffilmark{1}, H. Neyad \altaffilmark{2}, and M. A. Mojahedi \altaffilmark{3} } \affil{Department of
Physics, Semnan University, P. O. Box 35195-363, Iran}
\altaffiltext{1}{ hghafarnejad@ yahoo.com.}
\altaffiltext{2}{hsniad@yahoo.com.}
\altaffiltext{3}{amirmojahed2@gmail.com.}



\begin{abstract}
We obtain renormalized stress tensor of a mass-less, charge-less
dynamical quantum scalar field, minimally coupled with a spherically
symmetric static Lukewarm black hole. In two dimensional analog the
minimal coupling reduces to the conformal coupling and the stress
tensor is found to be determined by the nonlocal contribution of the
anomalous trace and some additional parameters in close relation to
the work presented by Christensen and Fulling. Lukewarm black holes
are a special class of Reissner- Nordstr\"{o}m-de Sitter space times
where its electric charge is equal to its mass. Having the obtained
renormalized stress tensor we attempt to obtain a time-independent
solution of the well known metric back reaction equation.
Mathematical derivations predict that the final state of an
evaporating quantum Lukewarm black hole reduces to a remnant stable
mini black hole with moved locations of the horizons. Namely the
perturbed black hole (cosmological) horizon is compressed (extended)
to scales which is smaller (larger) than the corresponding classical
radius of the event horizons. Hence there is not obtained an
deviation on the cosmic sensor-ship hypothesis.
\end{abstract}


\keywords{Hawking Radiation; Lukewarm Black hole; Back reaction
equation; Reissner Nordstr\"{o}m de Sitter; Noncommutative quantum
gravity; stability}



\section{Introduction}
Semiclassical approach of quantum gravity theory is known as quantum
 matter
 field theory propagated on a curved space-time,
  in which a classically treated curved space-time is perturbed by
 a suitable quantum matter field \citep{Bir82}.
  A fundamental problem in this version of the quantum gravity
  theory,
 is calculation of renormalized expectation value of quantum
 matter stress tensor operator $<\hat{T}_{\mu\nu}>_{ren}$. Renormalization theory give us a
suitable theoretical prediction, in which expectation value of a
singular quantum field stress tensor operator reduces to a
nonsingular quantity contained an anomalous trace. This nonsingular
stress tensor treats as source in RHS of the Einstein`s gravity
equation such as follows.
\begin{equation} G_{\mu\nu}-\Lambda
g_{\mu\nu}=8\pi \{T^{class}_{\mu\nu}+<\hat{T}_{\mu\nu}>_{ren}\}
\end{equation}
where $G_{\mu\nu}$ is Einstein tensor with the perturbed metric
$g_{\mu\nu}=\hat{g}_{\mu\nu}+\Delta g_{\mu\nu}$ and the background
metric $\hat{g}_{\mu\nu},$ $\Lambda$ is positive cosmological
constant and $T^{class}_{\mu\nu}$ is classical baryonic matter or
non-baryonic dark matter field stress tensor. Non-minimally coupled
scalar dark matter fields with a negative value of equation of state
parameter may to be come originally from effects of conformal
frames. The latter case of the matter is a good candidate to explain
positivity accelerated expansion of the universe and to remove the
naked singularity of the universe in quantum cosmological approach.
See \citep{Noz09} and references therein. The above equation which
is written in units $G=\hbar=c=1$ is called the metric back-reaction
equation. There are presented several methods for the
renormalization prescription, namely dimensional regularization,
point splitting, adiabatic and Hadamared renormalization
prescriptions \citep{Bir82}. The latter method has distinctions with
respect to the other methods of the renormalization prescriptions.
Hadamared renormalization prescription is described in terms of
Hadamared states and it predicts few conditions on unknown quantum
vacuum state of an arbitrary interacting quantum field
\citep{Bro84,Ber86,Gha97}. Hence it provides the most direct and
logical approach to the renormalization problem for practical
calculations. Furthermore it is well defined for both
massive and massless fields. \\
Renormalization theory is still establish the covariant conservation
of stress tensor operator expectation value of quantum field
contained with a non vanishing trace anomaly, namely
$\nabla^{\nu}<\hat{T}_{\mu\nu}>_{ren}=0.$  This anomaly is obtained
in terms of geometrical objects such as $R_{\mu\nu}R^{\mu\nu},$
$R_{\mu\nu\alpha\beta}R^{\mu\nu\alpha\beta},$ $\Box R,$ and $R^2$
for conformaly coupling massless quantum field propagated on four
dimensional curved space time
\citep{Chr76,Adl77,Wal78,Bir82,Bro84,Ber86,Gha97,Par09}.  In two
dimension the conformal coupling reduces to minimal coupling and so
the quantity of trace anomaly is obtained in terms of the Ricci
scalar $\mathcal{R}=R^{\beta}_{\beta}$ which for a massless scalar
matter field become: \begin{equation}
<\hat{T}_{\mu}^{\mu}>_{ren}=\frac{\mathcal{R}}{24\pi}.\end{equation}
The main problem in the equation (1) is to fined
$<\hat{T}_{\mu\nu}>_{ren}$ coupled with an arbitrary non-static and
non-spherically symmetric dynamical metric. But there are many
degrees of freedom and inherent complexity on four dimensional
solutions of equation (1). There are obtained in detail only for
class of four dimensional spherically symmetric space times which
are treated as two dimensional curved space times, because the
spherically symmetric condition on four dimensional space times
eliminates the extra degrees of freedom of Equation (1)
\citep{Chr77}. Two dimensional analog of the renormalization theory
and solutions of the back-reaction equation is used to determine
final state of spherically symmetric dilatonic and also
non-dilatonic evaporating black holes metric by several authors. For
instance Strominger et al were obtained a nonsingular metric for
final state of an evaporating two dimensional dilatonic massive
black hole \citep{Alw92,Ban92,Cal92,Rus92,Pir93}. It is shown in
\citep{Low93} that an evaporating two dimensional dilatonic Reissner
Nordestr\"{o}m
 black hole reduces to a remnant, stable nonsingular space time.
Evaporating dilatonic Schwarzschild de Sitter black holes final
state whose size is comparable to that of the cosmological horizon
is in thermal equilibrium \citep{Bou98}. It is obtained that final
state of a non-dilatonic Schwarzschild-de Sitter evaporating black
hole reduces to a remnant stable object with a nonsingular metric
 \citep{Gha06,Gha07}. It is shown by Balbinot et al that the Hawking evaporation \citep{Haw74,Haw75}
of the two dimensional non-dilatonic Schwarzschild black hole is
stopped \citep{Bbr84,Bal84,Bal85,Bal86,Bal89}. Back reaction
corrections of conformaly invariant quantum scalar
 field in the Hartle Hawking vacuum state
 \citep{Har76} was used to determine quantum perturbed metric of a non-dilatonic Reissner
Nordstr\"{o}m black hole by Wang et al \citep{Wan01}. They followed
the York approach where a small quantity $\epsilon$ is introduced to
solve the metric back-reaction equation (1) by applying the
perturbation method \citep{Yor85}.\\
 Furthermore noncommutative
quantum field theory in curved space times and so generalized
uncertainty principle derived from string theory \citep{Ama87,
Ama88, Ama89, Ama90,Cap00,Sny47,Sei99,Dou01}, is other quantum
gravity approach in which the space-time points might be
noncommutative \citep{Asc05,Cal05,Cal06,Cha01}. The latter quantum
gravity model is also predicts remnant stable mini-quantum black
hole where the Hawking radiation process finishes when black hole
approaches to
its Planck scale with a nonzero temperature \citep{Nic06, Noz05, Noz08}.\\
According to the perturbation method presented by the York, we solve
in this paper, two dimensional analog of the metric back-reaction
equation (1) and determine final state of an evaporating Lukewarm
black hole. This kind of a black hole is a special class of Reissner
Nordstr\"{o}m de sitter spherically symmetric static black hole
where mass parameter is equal to the charge parameter. According to
the work presented by Christensen and Fulling \citep{Chr77} we
obtain the renormalized stress tensor components of black hole
Hawking radiation
 in terms of a nonlocal contribution of the trace anomaly.
The plan of this paper is as follows.\\
 In section 2, we define Lukewarm
classical black hole metric and obtain locations of its event
horizons. In section 3,  we derive thermal radiation stress tensor
operator expectation value of a massless, charge-less quantum matter
scalar field propagating on the black hole metric. Having the
obtained Hawking radiation quantum stress tensor, we solve
back-reaction metric equation (1) in the section 4 and obtain
locations of the quantum perturbed horizons. Section 5 denotes to
the concluding remarks.

\section{Lukewarm Black Hole Metric}

%% In a manner similar to \objectname authors can provide links to dataset
%% hosted at participating data centers via the \dataset{} command.  The
%% second curly bracket argument is printed in the text while the first
%% parentheses argument serves as the valid data set identifier.  Large
%% lists of data set are best provided in a table (see Table 3 for an example).
%% Valid data set identifiers should be obtained from the data center that
%% is currently hosting the data.
%%
%% Note that AASTeX interprets everything between the curly braces in the
%% macro as regular text, so any special characters, e.g. "#" or "_," must be
%% preceded by a backslash. Otherwise, you will get a LaTeX error when you
%% compile your manuscript.  Special characters do not
%% need to be escaped in the optional, square-bracket argument.

Reissner Nordstr\"{o}m de Sitter space times with Lorentzian line
element is given by \begin{equation}
ds^2=-\Omega(r)dt^2+\frac{dr^2}{\Omega(r)}+r^2(d\theta^2+\sin^2\theta
d\varphi^2) \end{equation} where
\begin{equation}
\Omega(r)=1-\frac{2M}{r}+\frac{Q^2}{r^2}-\frac{\Lambda r^2}{3}
\end{equation}
 and $M,Q$ are the mass and charge of the black hole
respectively. $\Lambda$ is the positive cosmological constant.
Lukewarm black holes are a particular class of Reissner
Nordstr\"{o}m-de Sitter, with $Q=M.$ For $4M<\sqrt{3/\Lambda}$ we
have three distinct horizons, namely black hole event horizon at
$r=r_h,$ inner Cauchy horizon at $r=r_{ca},$ and cosmological
horizon at $r=r_c,$ where
\begin{equation}
r_{ca}=\frac{1}{2}\sqrt{3/\Lambda}\left(-1+\sqrt{1+4M\sqrt{\Lambda/3}}\right)
\end{equation}
\begin{equation}
r_h=\frac{1}{2}\sqrt{3/\Lambda}\left(1-\sqrt{1-4M\sqrt{\Lambda/3}}\right)
\end{equation}
 and
 \begin{equation}
r_c=\frac{1}{2}\sqrt{3/\Lambda}\left(1+\sqrt{1-4M\sqrt{\Lambda/3}}\right).
\end{equation} While the event horizon is formed by the gravitational
potential of the black hole, the cosmological horizon is formed as a
result of the expansion of the universe due to the cosmological
constant \citep{Gib77,Bre11}. An observer located between the two
horizons is causally isolated from the region within the event
horizon, as well as from the region outside the cosmological
horizon. The above line element is exterior metric of a spherically
symmetric static body with mass $M$ and charge $Q$. It is solution
of the equation (1) under the condition $<\hat{T}_{\mu\nu}>_{ren}=0$
where $T^{class}_{\mu\nu}$ is stress tensor of classical
electromagnetic field of a point charge $Q$ and it is given in
$(t,r,\theta,\varphi)$ coordinates such as follows:
\begin{equation}
{T^{(class)}}^{\mu}_{\nu}=\frac{1}{8\pi}\left(\frac{Q}{r^2}\right)^2\left(%
\begin{array}{cccc}
  1 & 0 & 0 & 0 \\
  0 &1 & 0 & 0 \\
  0 & 0 & -1 & 0 \\
  0 &0 & 0 & -1 \\
\end{array}%
\right).\end{equation} In advanced time Eddington-Finkelestein
coordinates $(v,r,\theta,\varphi)$ where
\begin{equation}
dv=dt+\frac{dr}{\Omega(r)} \end{equation} one can obtained classical
electromagnetic field stress tensor (8) such as follows.
\begin{equation}
T_{vv}^{class}(v,r)=\frac{\Omega^{-1}(x)-\Omega(x)}{128M^2x^4}
\end{equation} with
\begin{equation}
\Omega(x)=1-\frac{1}{x}+\frac{q^2}{4x^2}-\frac{\varepsilon x^2}{4},
\end{equation}
\begin{equation}
T^{class}_{vr}=T^{class}_{rr}=\frac{\Omega^{-1}(x)}{128M^2x^4}
\end{equation} and
\begin{equation}
 T^{class}_{\theta\theta}=-\frac{1}{32\pi
x^2},~~~T^{class}_{\varphi\varphi}=\sin^2\theta
T_{\theta\theta}^{class}
 \end{equation}where we defined
  \begin{equation}
x=\frac{r}{2M},~~~q=\frac{Q}{M},~~~\varepsilon=\frac{16M^2\Lambda}{3}>0.
\end{equation} Locations of the classical event horizons defined by
(5), (6) and (7) become respectively
\begin{equation}
x_{ca}=\frac{1-\sqrt{1+\sqrt{\varepsilon}}}{\sqrt{\varepsilon}},~~
~~x_b=\frac{1-\sqrt{1-\sqrt{\varepsilon}}}{\sqrt{\varepsilon}},
~~~x_c=\frac{1+\sqrt{1-\sqrt{\varepsilon}}}{\sqrt{\varepsilon}}
\end{equation} where
\begin{equation}
x_bx_c=\frac{1}{\sqrt{\varepsilon}} \end{equation}
 and in case
$0<\varepsilon<1$ we have
\begin{equation}
x_{ca}\approx\frac{\varepsilon}{8}-\frac{1}{2},~~~x_b\approx\frac{1}{2}+\frac{\sqrt{\varepsilon}}{8},~~
~x_c\approx\frac{2}{\sqrt{\varepsilon}}-\frac{1}{2}-\frac{\sqrt{\varepsilon}}{8}.
\end{equation} Applying (11) with $q=1,$ we obtain locations of the
horizons and quasi-flat regions of the black hole space time, from
the equations
 $\Omega(x)=0$ and $\frac{d\Omega(x)}{dx}=0$ respectively. These conditions reduce
to the following relations.
\begin{equation}
\varepsilon_e(x)=\frac{4}{x^2}-\frac{4}{x^3}+\frac{1}{x^4}.
\end{equation} and
\begin{equation}
\varepsilon_q(x)=\frac{2}{x^2}-\frac{1}{x^3}.\end{equation}
 Diagrams
of the functions defined by (18) and (19) are given by dash-lines
and solid line in figure 1, respectively. These diagrams are valid
for $0<\varepsilon<1.$ In case $\varepsilon\geq1$ locations of the
black hole and the cosmological horizons reach to each others and so
cases to instability of the black hole.\\
 In the next section we derive the
Hawking thermal radiation of a quantum Lukewarm black hole minimally
coupled with a linear two dimensional, massless, charge-less,
quantum scalar field. We will consider that the interacting quantum
scalar field to be charge-less and so has not electromagnetic action
with the classical electric field stress tensor $T^{class}_{\mu\nu}$
defined by (8). So we can suppose that the electric charge of the
black hole is not perturbed by the quantum scalar field. Also we
will assume that the quantum scalar field is propagated in s
(spherically) mode on the spherically symmetric background metric
(3) and so its $g_{tt}$ and $g_{rr}$ components are perturbed by the
renormalized expectation value of quantum field stress tensor
operator $<\hat{T}_{\mu\nu}[\hat{\phi}]>_{ren}.$ Applying the latter
assumption one can use two dimensional analog of the quantum field
back-reaction corrections on the metric such as follows.

%% In this section, we use  the \subsection command to set off
%% a subsection.  \footnote is used to insert a footnote to the text.


%% This section also includes several of the displayed math environments
%% mentioned in the Author Guide.

\section{Black Hole Hawking Radiation}
According to the work presented by Christensen and Fulling
\citep{Chr77} we will fined here general solution of the covariant
conservation equation defined by
\begin{equation}
\nabla_{\nu}S^{\nu}_{\mu}=0,~~~~S^{\nu}_{\mu}=<\hat{T}_{\mu}^{\nu}>_{ren}
\end{equation} under the anomaly condition (2). Assuming
$\theta,\varphi=constant,$ two dimensional analog of the metric (3)
described in the advanced time Eddington-Finkelestein coordinates
(9), become
\begin{equation}
ds^2=-\Omega(r)dv^2+2dvdr. \end{equation}
 Applying (21) the corresponding Ricci
scalar become $\mathcal{R}=\Omega^{\prime\prime}(r)$ where the over
prime $\prime$ denotes to differentiation with respect to radial
coordinate $r$ and hence the anomaly condition (2) become
\begin{equation}
S^{v}_{v}(r)+S^{r}_{r}(r)-\Omega^{\prime\prime}(r)/24\pi=0.
\end{equation} Nonzero components of second kind Christoffel symbols
are obtained as
\begin{equation}
\Gamma^v_{vv}=\frac{\Omega^{\prime}(r)}{2}
=-\Gamma^r_{vr}=\Gamma^r_{rv},~~~\Gamma^r_{vv}=\frac{\Omega(r)\Omega^{\prime}(r)}{2}.
\end{equation} Applying (23), the covariant conservation equation
defined by (20) leads to the following differential equations.
\begin{equation}
{S^{\prime}}^r_{v}+\Omega^{\prime}(S^r_{r}-S^v_v)/2-\Omega
\Omega^{\prime}S^v_{r}/2=0 \end{equation} and
\begin{equation}
{S^{\prime}}^r_r+\Omega^{\prime}S^v_r/2=0. \end{equation} Using
\begin{equation}
S^v_v=S_{rv},~~~~S^v_r=S_{rr},~~~S^r_r=S_{vr}+\Omega
S_{rr},~~~S^r_v=S_{vv}+\Omega S_{rv}
 \end{equation}with $S_{vr}=S_{rv}$ the
equations (22), (24) and (25) become respectively
\begin{equation} \Omega
S_{rr}+2S_{vr}=\frac{\Omega^{\prime\prime}}{24\pi},
 \end{equation}
 \begin{equation}
S_{vv}+\Omega S_{rv}=C_1
 \end{equation}and
 \begin{equation}
S_{vr}^{\prime}+\frac{3}{2}\Omega^{\prime}S_{rr}+\Omega
S^{\prime}_{rr}=0 \end{equation} where $C_1$ is integral constant.
Applying (27) and (29) we obtain
\begin{equation}
S_{rr}(r)=\frac{1}{\Omega^2(r)}\left\{C_2-\frac{1}{24\pi}
\int^{r}\Omega(\tilde{r})\Omega^{\prime\prime\prime}(\tilde{r})d\tilde{r}\right\}
\end{equation} where $C_2$ is also integral constant. Using (27) and
(30) one can show
\begin{equation} S_{vr}(r)=S_{rv}(r)=-\frac{C_2}{2\Omega(r)}
+\frac{1}{48\pi}\left\{\Omega^{\prime\prime}(r)+\frac{1}{\Omega(r)}\int^r\Omega
(\tilde{r})\Omega^{\prime\prime\prime}(\tilde{r})d\tilde{r}\right\}.
\end{equation}
 Applying (28) and (31) we obtain
\begin{equation}  S_{vv}(r)=C_1+\frac{C_2}{2}-\frac{1}{48\pi}\left\{\Omega(r)\Omega^{\prime\prime}(r)+
 \int^{r}\Omega(\tilde{r})\Omega^{\prime\prime\prime}(\tilde{r})d\tilde{r}\right\}.
\end{equation}
  Using (4) and (14) with $q=1,$ $0<\varepsilon<1,$ the
stress tensor components defined by (30), (31) and (32) can be
rewritten as
\begin{equation}
S_{\mu\nu}(v,r)=\frac{1}{96\pi M^2}$$$$\times\left(%
\begin{array}{cc}
48\pi M^2(2C_1+C_2)-2B(x)-12A(x) & \frac{2B(x)+12A(x)-48\pi M^2C_2}{\Omega(x)} \\
\frac{2B(x)+12A(x)-48\pi M^2C_2}{\Omega(x)} &  \frac{96\pi M^2C_2-12A(x)}{\Omega^2(x)}\\
\end{array}%
\right)\end{equation} where we defined
\begin{equation}
\Omega(x)=1-\frac{1}{x}+\frac{1}{4x^2}-\frac{\varepsilon x^2}{4},
\end{equation}
\begin{equation}
A(x)=\frac{1}{6}\int^x\Omega(\tilde{x})\Omega^{\prime\prime\prime}
(\tilde{x})d\tilde{x}=\frac{1}{24x^6}-\frac{1}{4x^5}+\frac{1}
{2x^4}-\frac{1}{3x^3}-\frac{\varepsilon}{8x^2}+\frac{\varepsilon}{4x}
\end{equation} and
\begin{equation}
\Omega(x)\Omega^{\prime\prime}(x)=B(x)=\frac{3}{8x^6}
-\frac{2}{x^5}+\frac{7}{2x^4}-\frac{2}{x^3}
-\frac{\varepsilon}{2x^2}+\frac{\varepsilon}{x}-\frac{\varepsilon}{2}+\frac{\varepsilon^2
x^2}{8}.
 \end{equation} Now we should be determine the integral constants
$C_{1}$ and $C_2.$ For the determination of these constants we
require the regularity of $S^{\mu}_{\nu}$ at the black hole horizon
in a coordinate system which is regular there. The stress tensor
$S^{\mu}_{\nu},$ as measured in a local Kruskal coordinate system at
black hole horizon, will be finite if $S_{vv}$ and $S^t_t+S^r_r,$
are finite as $x\to x_b$ and
\begin{equation}\lim_{x\to
x_b}(x-x_b)^{-2}|S_{uu}|<\infty, \end{equation} where $(u,v)$ are
null coordinates \citep{Chr77}. We find easily
\begin{equation}
S_{uu}=\frac{1}{4}(S_{tt}+\Omega^2S_{rr}-2\Omega S_{tr})
\end{equation} where
\begin{equation} S_{tt}=S_{vv}+\Omega^2S_{rr}-2\Omega
S_{rv} \end{equation} and
\begin{equation} S_{tr}=S_{rt}=S_{vr}-\Omega S_{rr}
\end{equation} are obtained by applying (9) and definition
$S_{\mu\nu}=\delta_{\mu}^{\alpha}\delta_{\nu}^{\beta}
S_{\alpha\beta}.$ Applying (30), (38),  (39) and (40) we obtain
\begin{equation}S_{uu}(x)=156\pi M^2C_2+24\pi M^2C_1-27A(x)-5B(x)/2.
\end{equation} For a fixed $\varepsilon$ as $0<\varepsilon<1,$
diagram of the figure 1 determines locations of the unperturbed
black hole and cosmological horizons $x_b,x_c$  where $x_b<x_c.$
Having this obtained black hole horizon radius $x_b,$ and (41), the
initial condition $S_{uu}(x_b)=0$ reduces to
\begin{equation}
2C_1+13C_2=\frac{54A(x_b)+5B(x_b)}{24\pi M^2}. \end{equation} For
fields describing a gas of massless bosons (without spin, charge, or
internal degrees of freedom) moved in quasi flat regions of a two
dimensional curved space, the density and the flux are actually
equal, so that $S^t_t(x_q)+S^r_r(x_q)=0,$ \citep{Chr77} which in
terms of the $(v,r)$
  coordinates become
  \begin{equation} 2\Omega(x_q)S_{rv}(x_q)-S_{vv}(x_q)=0
  \end{equation}
where $x_q$ obtained from $\Omega^{\prime}(x_q)=0,$ (see figure 1 )
defines quasi-flat regions of two dimensional version of the space
time (3). Applying (33) the initial condition (43) become
\begin{equation}
2C_1+3C_2=\frac{18A(x_c)+3B(x_c)}{24\pi M^2}. \end{equation} Using
(42) and (44) one can obtain
\begin{equation}
C_{1}=\frac{18[13A(x_c)-9A(x_b)]+13B(x_c)-15B(x_b)}{480 \pi M^2}
\end{equation} and
\begin{equation}
C_2=\frac{18[3A(x_b)-A(x_c)]+5B(x_b)-B(x_c)}{240\pi M^2}.
\end{equation} We are now in a position to show that the stress
tensor (33) defined in the quasi flat region $x=x_q$ can be
decomposed in terms of thermal equilibrium ${S^{(e)}}^{\nu}_{\mu}$
and radiating ${S^{(r)}}^{\nu}_{\mu}$ stress energy tensors of
massless and charge-less bosonic gas
 respectively as
 \begin{equation} {S^{(e)}}_{\nu\mu}(t,r)=\frac{\pi}{12}T_c^2\left(%
\begin{array}{cc}
  -2 & 0 \\
  0 & 2 \\
\end{array}%
\right)
\end{equation}and
\begin{equation} {S^{(r)}}_{\nu\mu}(t,r)= \frac{\pi}{12}T_b^2\left(%
\begin{array}{cc}
  -1 & 1 \\
  1 & 1 \\
\end{array}%
\right) \end{equation}where
\begin{equation}
\frac{T_b}{T_{S}}=4\sqrt{B(x_q)+12A(x_q)-16.2A(x_b)+6.4A(x_c)-3.9B(x_c)+4.5B(x_b)},
\end{equation} and
\begin{equation}
\frac{T_c}{T_S}=2\sqrt{54A(x_b)-18A(x_c)+5B(x_b)-B(x_c)-2B(x_q)-24A(x_q)}
\end{equation} are defined as the black hole radiation and the
cosmological thermal equilibrium temperatures respectively.
$T_{S}=\frac{1}{8\pi M}$ is the well known Schwarzschild black hole
temperature. Now we seek to obtain time-independent solutions of the
back reaction equation (1) by applying (10), (11), (12) and (33) in
case $q=1.$

%% The equation environment wil produce a numbered display equation.

%% The \notetoeditor{TEXT} command allows the author to communicate
%% information to the copy editor.  This information will appear as a
%% footnote on the printed copy for the manuscript style file.  Nothing will
%% appear on the printed copy if the preprint or
%% preprint2 style files are used.

%% The eqnarray environment produces multi-line display math. The end of
%% each line is marked with a \\. Lines will be numbered unless the \\
%% is preceded by a \nonumber command.
%% Alignment points are marked by ampersands (&). There should be two
%% ampersands (&) per line.

%% Putting eqnarrays or equations inside the mathletters environment groups
%% the enclosed equations by letter. For instance, the eqnarray below, instead
%% of being numbered, say, (4) and (5), would be numbered (4a) and (4b).
%% LaTeX the paper and look at the output to see the results.

%% This section contains more display math examples, including unnumbered
%% equations (displaymath environment). The last paragraph includes some
%% examples of in-line math featuring a couple of the AASTeX symbol macros.

\section{Back Reaction Equation}

%% The displaymath environment will produce the same sort of equation as
%% the equation environment, except that the equation will not be numbered
%% by LaTeX.
Applying the advanced-time Eddington-Finkelestein coordinates
$(v,r,\theta,\varphi),$ defined by (9), the quantum perturbed metric
(3) is taken to have the form \begin{equation}
ds^2_f=-e^{2\psi(r)}F(r)dv^2+2e^{\psi(r)}dvdr+r^2(d\theta^2+\sin^2\theta
d\varphi^2) \end{equation} with
\begin{equation}
F(r)=1-\frac{2m(r)}{r}+\frac{Q^2}{r^2}-\frac{1}{3}\lambda(r)r^2
\end{equation} in which $\psi,m$ are assumed to be depended alone to
the radial coordinate $r,$ because the perturbed metric should still
be static and spherically symmetric. The index $f$ denotes to the
word $final$ state of quantum perturbed evaporating Lukewarm Black
hole. The perturbed metric (51) leads to the static metric (3) under
the following boundary conditions:
\begin{equation}
\psi(x_b;\varepsilon=0)=0,~~~m(x_b;\varepsilon=0)=M,~~~~\lambda(x_b;\varepsilon=0)=\Lambda
\end{equation} where $x_b=\frac{1}{2}$ is obtained from (15) under
the condition $\varepsilon=0.$ Applying (51) and definitions

\begin{equation}\frac{m(r)}{M}=\rho(x),~~~\lambda(x)=
\frac{3\varepsilon\sigma(x)}{16M^2},~~~q=1=\frac{Q}{M},~~~x=\frac{r}{2M}
\end{equation} the $(v,r)$ components of the Einstein`s tensor become

\begin{equation}
G_{vv}(x)=-\frac{e^{2\psi(x)}}{x^2}\left(1-\frac{\rho(x)}{x}+\frac{1}{4x^2}-\frac{\varepsilon\sigma(x)
x^2}{4}\right)$$$$\times\left(\rho^{\prime}(x)+\frac{1}{4x^2}+\frac{3\varepsilon\sigma(x)x^2}{4}+
\frac{\varepsilon\sigma^{\prime}(x)x^3}{4}\right), \end{equation}

\begin{equation} G_{vr}(x)=G_{rv}(x)=e^{\psi(x)}
\left(\frac{\rho^{\prime}(x)}{x^2}+\frac{\varepsilon\sigma^{\prime}(x)x}{4}
+\frac{1}{4x^4}+\frac{3\varepsilon\sigma(x)}{4 }\right)
\end{equation}
\begin{equation} G_{rr}(x)=-2\frac{\psi^{\prime}}{x}
\end{equation}where
 $\prime$ denotes to
differentiation with respect to $x$. All other components are zero
except $G_{\theta}^{\theta}=G_{\varphi}^{\varphi}$ which follows
from the Binachi identity $\nabla_{\xi}G^{\xi}_r=0.$ Applying (10),
(11), (12), (33), (55), (56) and (57), we obtain $vv,$ $vr$ and $rr$
components of the Back-reaction equation (1) as respectively

 \begin{equation}
  \Omega(x) e^{2\psi(x)}\left(1-\frac{\rho(x)}{x}+\frac{1}
  {4x^2}-\frac{\varepsilon\sigma(x)x^2}{4}\right)\left(\frac{1}{16x^4}+
 \frac{\rho^{\prime}(x)}{4x^2}+\frac{\varepsilon\sigma^{\prime}(x)x}{16}
 \right)$$$$+\frac{\pi[1-\Omega^2(x)]}{16x^2}+\Omega(x)[4\pi M^2(2C_1+C_2)-A(x)-B(x)/6]=0
 \end{equation}
 \begin{equation}
\Omega(x) e^{\psi(x)}\left(\frac{1}{16x^4}+
 \frac{\rho^{\prime}(x)}{4x^2}+\frac{\varepsilon\sigma^{\prime}(x)x}{16}
 \right)+4\pi M^2C_2-\frac{\pi}{16x^2}-A(x)-\frac{B(x)}{6}=0,
 \end{equation} and
 \begin{equation}
\psi^{\prime}(x)=\frac{16x^4[A(x)-8\pi
M^2C_2]-\pi\Omega(x)}{8x^3\Omega^2(x)} \end{equation} where
$\Omega(x)$ is given by (34). Solution of the equation (60) can be
obtained directly by integrating. It is useful that, we obtain
behavior of the solution $\psi(x)$ at neighborhood of its singular
points, namely $x=0$ and $x_{b,c}$ where $\Omega(x_{b,c})=0.$
However we obtain

\begin{equation} \psi(x<x_b)\simeq
C_{\psi}+0.24\ln \left(4-\frac{1}{x}\right)+\frac{0.3125}{4x-1},
\end{equation}

  \begin{equation} \psi(x\to x_b)\simeq C_{\psi}-\frac{2x_b^3[A(x_b)-8\pi M^2C_2]}{(x-x_b)}
  \end{equation}
  and
  \begin{equation} \psi(x\to x_c)\simeq C_{\psi}+\frac{x_c^3[A(x_c)-8\pi M^2C_2]}{2(x_c-x)}
  \end{equation}
  where
  \begin{equation}\Omega(x<x_b)\simeq\frac{1}{4x^2}-\frac{1}{x},~~~0<\varepsilon<1,
  \end{equation}

 \begin{equation} \Omega(x\to x_b)\simeq1-\frac{x_b}{x},~~
 ~\Omega(x\to x_c)\simeq1-\frac{x^2}{x_c^2}\simeq2(1-\frac{x}{x_c})
 \end{equation}
and $C_{\psi}$ is integral constant which is determined by the
initial conditions (53) such as follows.\\
 Applying $\psi(x_b)=0$
where $x_b=\frac{1}{2}$ with $\varepsilon=0$ the solution (61) leads
to
\begin{equation}
C_{\psi}\simeq2.07\times10^{-3},~~~e^{C_{\psi}}\approx1.
\end{equation} Inserting (59) the equation (58) become
\begin{equation}
\frac{\rho(x)}{x}+\frac{\varepsilon
x^2\sigma(x)}{4}=\frac{H(x)}{G(x)} \end{equation} where
\begin{equation}
H(x)=(1+4x^2)[\pi/4x^2+4[A(x)+B(x)/6]-16\pi M^2C_2]$$$$+\{\pi+64\pi
M^2 x^2(2C_1-C_2)\Omega(x)-[\pi+16
x^2(A(x)+B(x)/6)]\Omega^2(x)\}e^{-\psi(x)} \end{equation} and
\begin{equation}
G(x)=\pi-64\pi M^2C_2 x^2+16x^2[A(x)+B(x)/6]. \end{equation} One can
rewrite the equation (59) as
\begin{equation}
\frac{\rho^{\prime}(x)}{x^2}+\frac{\varepsilon
x\sigma^{\prime}(x)}{4}=Z(x) \end{equation} where we defined
\begin{equation}
Z(x)=\frac{\pi+16x^4[A(x)+B(x)/6]-64\pi M^2C_2 x^4-\Omega(x)
e^{\psi(x)}}{4x^4\Omega(x) e^{\psi(x)}}. \end{equation} Applying
(67), (70) and identity
\begin{equation}
\frac{2\rho(x)}{x^3}-\frac{\varepsilon\sigma(x)}{4}=\frac{\rho^{\prime}}{x^2}+\frac{\varepsilon
x\sigma^{\prime}}{4}-\left[\frac{1}{x}\left(\frac{\rho(x)}{x}+\frac{\varepsilon\sigma(x)
x^2}{4}\right)\right]^{\prime} \end{equation} we obtain
\begin{equation}
\frac{2\rho(x)}{x^3}-\frac{\varepsilon\sigma(x)}{4}=Z(x)-\left[\frac{H(x)}{xG(x)}\right]^{\prime}.
\end{equation} Using (67) and (73) we obtain exactly

\begin{equation}\rho(x)=\frac{x^3}{3}\left[Z(x)+\frac{1}{x}\left(\frac{H(x)}{G(x)}\right)^{\prime}\right]
\end{equation} and
\begin{equation}
\sigma(x)=\frac{4}{3\varepsilon}\left[-Z(x)+\frac{1}{x}
\left(\frac{H(x)}{G(x)}\right)^{\prime}+\frac{H(x)}{x^2G(x)}\right].
\end{equation} Applying (35), (36), (61), (62), (63),
(64), (65), and (66) one obtain
\begin{equation}
H(x<x_b)\simeq\frac{0.42}{x^6}\left(1-\frac{1}{x^{1.76}}\right),
~~~G(x<x_b)\simeq\frac{9.3}{x^3}\left(\frac{0.18}{x}-1\right),
\end{equation}
\begin{equation} H(x\to x_b)\simeq\pi
\exp\left\{\frac{2x_b^3[A(x_b)-8\pi M^2C_2]}{(x-x_b)}\right\}
\end{equation}
\begin{equation} H(x\to
x_c)\simeq(1+4x_c^2)[\pi/4x_c^2+4A(x_c)+2B(x_c)/3-16\pi
M^2C_2]$$$$\times \pi\exp\left\{-\frac{x_c^3[A(x_c)-8\pi
M^2C_2]}{2(x_c-x)}\right\} \end{equation}

   \begin{equation} G(x\to x_{b,c})=\pi-64\pi M^2C_2
x_{b,c}^2+16x_{b,c}^2[A(x_{b,c})+B(x_{b,c})/6], \end{equation}

\begin{equation}
Z(x<x_b)\simeq\frac{2.27}{x^3}\left(1-\frac{0.11}{x}\right),
\end{equation}
\begin{equation}
Z(x\to
x_b)\simeq\left[\frac{\pi}{4x_b^3}+4x_b[A(x_b)+B(x_b)/6]-16\pi M^2
C_2
x_b\right]$$$$\times(x-x_b)^{-1}\exp\left\{\frac{2x_b^3[A(x_b)-8\pi
M^2C_2]}{(x-x_b)}\right\} \end{equation} and
\begin{equation} Z(x\to
x_c)\simeq\left[\frac{\pi}{8x_c^3}+2x_c[A(x_c)+B(x_c)/6]-8\pi
M^2C_2x_c\right]$$$$\times(x_c-x)^{-1}\exp\{-\frac{x_c^3[A(x_c)-8\pi
M^2C_2]}{2(x_c-x)}\} \end{equation} Using (76) and (80), the
equations defined by (74) and (75) become respectively
\begin{equation}
\rho(x<x_b)\simeq0.76\left(1-\frac{0.11}{x}\right)-\frac{2.51\times10^{-3}}{x^{5.52}(0.18-x)^2}
\end{equation} and
\begin{equation}
\sigma(x<x_b)\simeq-\frac{4}{3\varepsilon}\left\{\frac{2.27}{x^{3}}
\left(1-\frac{0.11}{x}\right)+\frac{0.043}{x^{5.76}(0.18-x)}+\frac{7.53\times10^{-5}}{x^{8.52}(0.18-x)^2}\right\}.
\end{equation} Applying (77), (79) and (81) the equations
defined by (73) and (75) become respectively
\begin{equation}\rho(x\to
x_b)\simeq\{\frac{\frac{\pi}{12}+\frac{4x_b^4[A(x_b)+B(x_b)/6]}{3}-\frac{16\pi
M^2C_2x_b^4}{3}}{(x-x_b)}$$$$ -\frac{\frac{2\pi x_b^5[A(x_b)-8\pi
M^2C_2]}{3G(x_b)}}{(x-x_b)^{2}}\}\exp\left\{\frac{2x_b^3[A(x_b)-8\pi
M^2C_2]}{(x-x_b)}\right\}, \end{equation}
\begin{equation}
\sigma(x\to
x_b)\simeq-\frac{4}{3\varepsilon}\{\frac{\pi/3x_b^3+4x_b[A(x_b)+B(x_b)/6]-16\pi
M^2 x_bC_2}{(x-x_b)}$$$$+\frac{2\pi x_b^2[A(x_b)-8\pi
M^2C_2]}{G(x_b)(x-x_b)^{2}}\}\exp\left\{\frac{2x_b^3[A(x_b)-8\pi
M^2C_2]}{(x-x_b)}\right\}, \end{equation}
\begin{equation} \rho(x\to
x_c)\simeq\{\frac{\pi/24+2x_c^4[A(x_c)+B(x_c)/6]/3-8\pi
M^2C_2x_c^4/3}{x_c-x}$$$$-\frac{\pi
x_c^3(1+4x_c^2)[\pi/4+4A(x_c)+2B(x_c)/3-16\pi M^2C_2][A(x_c)-8\pi
M^2C_2]}{2G(x_c)(x_c-x)^2}\}$$$$\times\exp\left\{-\frac{x_c^3[A(x_c)-8\pi
M^2C_2]}{2(x_c-x)} \right\} \end{equation} and

 \begin{equation}
  \sigma(x\to x_c)\simeq -\frac{4}{3\varepsilon}\{\frac{\pi/8x_c^3+2x_c[A(x_c)+B(x_c)/6]-8\pi
M^2C_2x_c}{x_c-x}$$$$+\frac{\pi
x_c^2(1+4x_c^2)[\pi/4x_c^2+4A(x_c)+2B(x_c)/3-16\pi
M^2C_2][A(x_c)-8\pi
M^2C_2]}{2G(x_c)(x_c-x)^2}\}$$$$\times\exp\left\{-\frac{x_c^3[A(x_c)-8\pi
M^2C_2]}{2(x_c-x)} \right\} \end{equation} Having the above obtained
solutions we are now in a position to write the quantum perturbed
Lukewarme black hole metric (51) defined in $(t,r,\theta,\varphi)$
coordinates as
\begin{equation}
ds^2_f=-F(r)dt^2+\frac{dr^2}{F(r)}+r^{2}\{d\theta^2+\sin^2\theta
d\varphi^2\} \end{equation}in which
\begin{equation}
dt=e^{\psi(r)}dv-\frac{dr}{F(r)} \end{equation} and $\psi(r)$ with
$r=2Mx,$ is given by (61), (62) and (63). $F(r)$ defined by (52) and
(54) as
\begin{equation}
F(x)=1+\frac{1}{4x^2}-\frac{\rho(x)}{x}-\frac{\varepsilon\sigma(x)x^2}{4}
\end{equation} is given exactly by applying (83), (84), (85),
(86), (87) and (88). It will be useful that we choose a
numerical value for $x_{b,c}$ from the figure 1 such as follows.\\
Experimental limits on the cosmological constant is obtained as
\citep{Ken91}
\begin{equation}
 |\Lambda|\leq10^{-54}cm^{-2}
 \end{equation} and order of magnitude of
Schwarzschild radiuses for a galaxy and the Sun is given
 by $(2M\sim10^{16} cm)$ and $(2M\sim3\times10^5 cm)$ respectively. So
whose corresponding coupling parameter
$\varepsilon=\frac{16M^2\Lambda}{3}$ will be obtain as
$\varepsilon_{galaxy}\simeq1.33\times10^{-22}$ and
$\varepsilon_{sun}\simeq1.2\times10^{-43}$  respectively which are
very small digits. As a numerical result we use here
$\varepsilon=10^{-22}$ and obtain
\begin{equation} (x_b,x_c)\cong(0.5,10^{11})
\end{equation}
\begin{equation}
A(x_b)=A(x_c)\cong0~~~B(x_b)\cong48,~~~B(x_c)\cong-3.75\times10^{-23}
\end{equation} and
 \begin{equation}\pi
M^2C_2\cong1,~~~G(x_b)\cong-15,~~~G(x_c)\cong12.57. \end{equation}
Using (83), (84), (85), (86), (87), (88) and the above numerical
values the equation (91) leads to
\begin{equation}
F(x<0.5)\cong\frac{0.62}{x^{6.52}(0.18-x)^2}, \end{equation}
\begin{equation}
F(x\to
0.5)\cong\frac{42.53(x-0.45)}{(x-0.5)^2}\exp\left\{\frac{2}{0.5-x}\right\},
\end{equation}
\begin{equation} F(x\to
10^{11})\cong\frac{2.13\times10^{56}}
{\left(1-\frac{x}{10^{11}}\right)^2}\exp\left\{\frac{4\times10^{12}}{1-\frac{x}{10^{11}}}\right\}.
\end{equation} The solution (96) dose not vanished in regions
$0<x<0.5.$ The solution (97) vanishes at $x\cong0.45.$ This is
location of the perturbed black hole event horizon where $x_b=0.5$
is classical unperturbed radius of the Lukewarm black hole event
horizon. It is seen easily that the solution (98) converges to a
zero value (the perturbed cosmological event horizon) at limits
$x>>10^{11}.$ These solutions predict that the interacting quantum
field back reaction corrections on the perturbed Lukewarm static
black hole metric cause to shift the location of event horizons. In
other word the cosmic sensor-ship hypothesis is still saved in the
presence of the quantum field perturbations on a curved background
metric. As a future work the authors will be attempt to seek a time
dependent version of perturbation solutions of the problem.
Particularly stability prediction of an evaporating Lukewarm black
hole encourages us to seek unperturbed solutions of the back
reaction equation of the problem by using the Wheeler-DeWitt
canonical quantum gravity approach. Result of this work together
with results of several works pointed in the introduction predict
remnant stable mini quantum black holes where the cosmic sensor-ship
hypothesis is still valid.
\section{Concluding Remarks}
 Two dimensional analog of the Hawking thermal radiation stress tensor
of the quantum perturbed spherically symmetric static Lukewarm back
hole is derived, by applying the Christensen and Fulling method.
Then the obtained stress tensor, is used to solve a time-independent
version of the well known metric back-reaction equation defined in a
perturbed Lukewarm metric. According to the York`s hypothesis
\citep{Yor85}, we assume here that the massless and charge-less
quantum scalar fields propagated on the background metric are in s
(spherically) modes and so $(t,r)$ components of the metric are
perturbed only. This leads still to save its spherically symmetric
property and to assume that the mass and cosmological parameter of
the Lukewarm black hole to be chosen as slowly varying radial
dependence functions. However, mathematical derivations predict a
shrunk black hole horizon with an extended cosmological horizon
 with respect to the corresponding classical horizons location.
 Particularly these quantum field perturbations do not cause
 violations of the cosmic sensor-ship hypothesis.

\begin{figure}
\epsscale{.80} \plotone{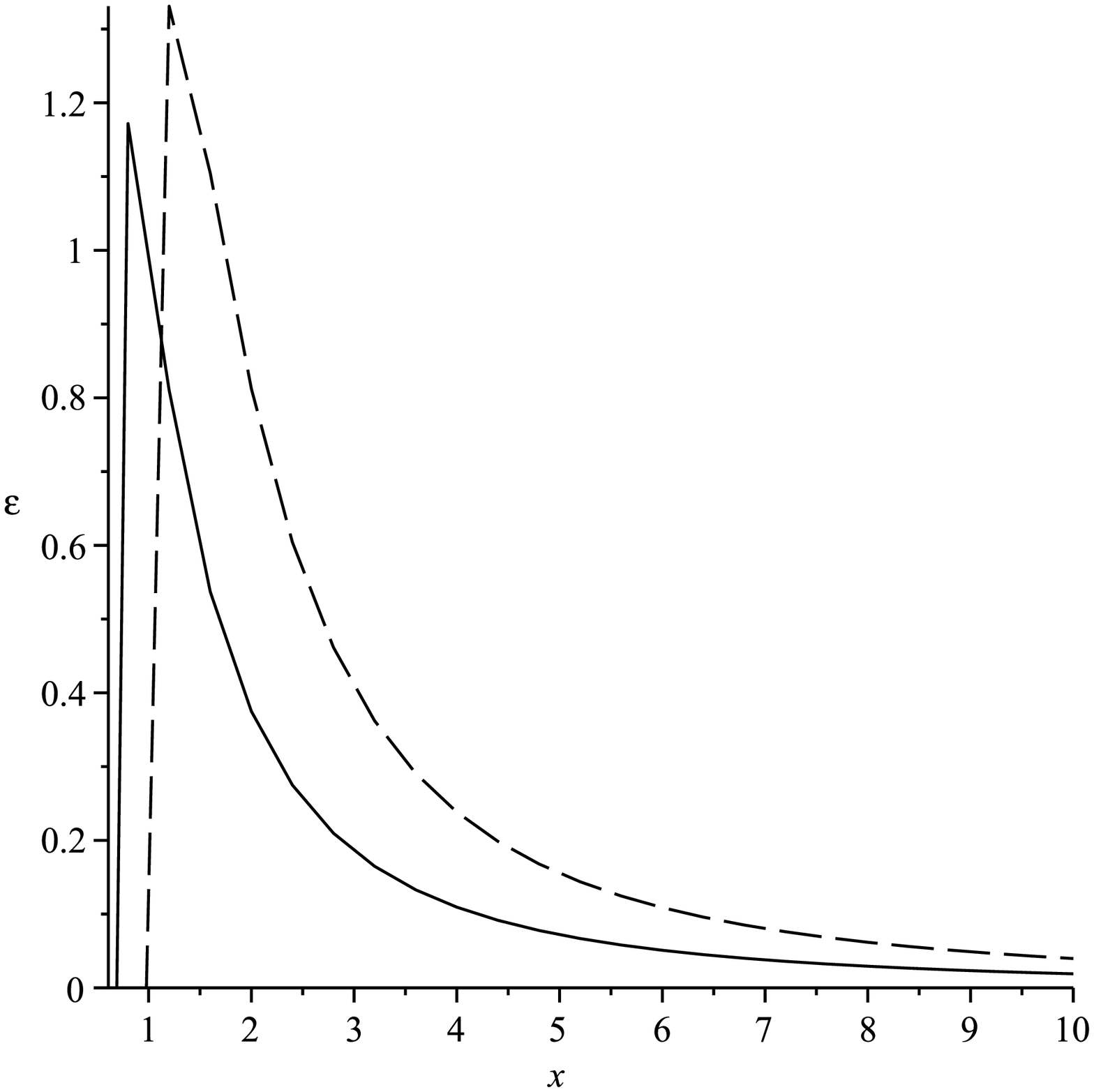} \caption{Dash-lines describe the
black hole and cosmological horizon radiuses obtained from the
equation $\Omega(x)=0$ with $q=1,$ namely the equation (18). Solid
line defines quasi flat regions of the space time (3) which is
obtained from the equation $\Omega^{\prime}(x)=0$ with $q=1,$ namely
the equation (19). Values with $0<\varepsilon<10^{-22}$ is not shown
here, because the diagram has large variations. \label{fig1}}
\end{figure}

\end{document}